\documentclass[12pt,a4paper]{article}
\usepackage[latin1]{inputenc}
\usepackage[T1]{fontenc}
\usepackage{amsmath}
\usepackage{amsfonts}
\usepackage{amssymb}
\usepackage[english]{babel}
\usepackage[dvips]{graphicx,color}
\newcommand{\ecua}{\begin{equation}}
\newcommand{\fin}{\end{equation}}
\newcommand{\im}{\dot\imath}
\newcommand{\dmu}{\partial_{\mu}}

\newcommand{\fii}{\varphi}
\newcommand{\fiid}{\varphi^{*}}
\newcommand{\punto}{\;\; .}
\newcommand{\coma}{\;\; ,}
 
\def\bra#1{\mathinner{\langle{#1}|}} 
\def\ket#1{\mathinner{|{#1}\rangle}} 
\def\braket#1{\mathinner{\langle{#1}\rangle}}

{\catcode`\|=\active\gdef\Braket#1{\left<\mathcode`\|"8000\let|\bravert {#1}\right>}} 
\def\bravert{\egroup\,\vrule\,\bgroup}
\title{One-loop effects in a self-dual planar noncommutative theory}
\author{C.~D.~Fosco,
and
G.~A.~Moreno\\
{\normalsize\it Centro At\'omico Bariloche and Instituto Balseiro}\\
{\normalsize\it Comisi\'on Nacional de Energ\'\i a At\'omica}
\\
{\normalsize\it R8402AGP Bariloche, Argentina.}}
\begin{document}
\date{\today}
\maketitle
\begin{abstract}
We study the UV properties, and derive the explicit form of the one-loop
effective action, for a noncommutative complex scalar field theory in $2+1$
dimensions with a Grosse-Wulkenhaar term, at the self-dual point.  We also
consider quantum effects around non-trivial minima of the classical action
which appear when the potential allows for the spontaneous breaking of the
$U(1)$ symmetry. For those solutions, we show that the one-loop correction
to the vacuum energy is a function of a special combination of the
amplitude of the classical solution and the coupling constant.
\end{abstract}
\section{Introduction}\label{sec:intro}
Noncommutative Quantum Field Theories (NCQFT's), in particular those
obtained by Moyal deformation of the usual (pointwise) product of
functions, have been a subject of intense research in recent
years~\cite{nekrasov}, because of many different reasons. Among them is
their relevance to open string dynamics~\cite{SW} and, in a quite different
context, they are important tools for an effective description of the
Quantum Hall Effect (QHE)~\cite{Suss}. In this  realization of an
incompressible quantum fluid~\cite{Jackiw:2004nm}, the projection to the
lowest Landau level under the existence of a strong magnetic field amounts,
for a two-dimensional system, to the noncommutativity of the spatial
coordinates~\cite{Duval:2000xr}.

In this paper, we calculate one-loop quantum effects around both trivial
and non-trivial saddle points,  for the NCQFT of a self-interacting complex
scalar field equipped with a Grosse-Wulkenhaar (GW) term~\cite{bulquen}
(see also~\cite{R} and \cite{0710.2652}).  

One of the interests for carrying out this explicit calculation is that,
in spite of the many important general results for this kind of
NCQFT~\cite{bulquen} there are, we believe, still few concrete results
obtained by actually evaluating quantum effects in models that
include a GW-term. In particular, we shall focus on the divergent terms in
the effective action, and on the first quantum corrections to the effective
action around non trivial minima, in the case of a spontaneous symmetry
breaking potential. 

We deal with a $2+1$ dimensional model, something which makes it 
more attractive from the point of view of its potential applications to the
situation of a planar system in an external magnetic field. At the same
time, it provides an opportunity to probe the effect to the GW term in an
odd number of spacetime dimensions where, necessarily, some of the
coordinates do commute. Finally, we also consider the important issue of
calculating quantum corrections on top of non-trivial minima that arise
when there is spontaneous symmetry breaking. 

This article is organized as follows: in section~\ref{sec:themodel} we
write down the action that defines the model, selecting the basis of
functions to be used in the loopwise expansion, and extracting the
resulting Feynman rules. We analyze the renormalizability 
of the theory in~\ref{sec:renormalization}, while 
the one-loop corrections to the two and four-point functions are evaluated in
section~\ref{sec:renormalizedgeneratingfunctional}. We consider quantum
effects around non-trivial minima in section~\ref{ntvc}.  In
section~\ref{sec:conclusions}, we present our conclusions.

\section{The model}\label{sec:themodel} 
We are concerned with a noncommutative model whose dynamical variable is a
complex scalar field in $2+1$ space-time dimensions, such that the
coordinates satisfy: 
\begin{equation}
[x_\mu \,,\, x_\nu ] \;=\; i \, \theta_{\mu\nu} 
\;\;, \;\;\; \mu, \, \nu \,=\, 0,\,1,\,2 \;,
\end{equation} 
where $\theta_{\mu\nu}$ are the elements of a constant real antisymmetric 
constant matrix.
In $2+1$ dimensions this matrix is necessarily singular; thus we shall
assume that its (only) null eigenvalue corresponds to the time direction,
$x_0$, since we are not interested in introducing noncommutativity for the
time coordinate. Although there are some general arguments to discard those
theories~\cite{alvarezgaume}, in our case the reason is simpler: we want to
consider theories that might be interpreted in terms of effective field
theory models in strong magnetic fields~\cite{cesar_ana_lopez}. Thus we 
have the more explicit commutation relations: 
\begin{equation} 
[ x_0, x_j ] \,=\, 0 \;\;,\;\;\;\; [ x_j, x_k ] \,=\,i \,
  \theta_{jk} \;\;\;,\;\;\; j,\,k = 1,\, 2 \;
\end{equation}
where $\theta_{jk}= \theta \,\epsilon_{jk}$, and we shall assume that 
$\theta > 0$.

The model is defined by the following Euclidean action: 
\ecua
\mathcal{S}=\int_{x,t} \left(\dmu \fiid \dmu \fii+m^{2} \fiid \fii+\Omega^2
  \fiid \star z_j \star \fii \star z_j \right)\,+\,\mathcal{S}_{int} 
\fin
(with $z_j\equiv\theta^{-1}_{jk} x_k$), which is of the kind proposed in~\cite{bulquen}.
Under the extra assumption that $\Omega^2\equiv 2$, the system is said to be
at the self-dual point since it is invariant under a combined Fourier
transformation and rescaling~\cite{szabo} of spatial coordinates:
\ecua
\mathcal{S}[\fii,\fiid,\theta,g]=\mathcal{S}[\frac{1}{\theta}\hat \fii_{(\frac{x}{\theta})},\frac{1}{\theta}\hat \fiid_{(\frac{x}{\theta})},\theta,g] \;\; ,
\fin
a symmetry that survives, as we shall see explicitly, one loop quantum
corrections.
At that special point, one of the terms in the commutators used to define
the spatial (inner) derivatives is canceled with a like one coming from
the confining potential term, leading to an action with the form:
\ecua \label{eq: accion}
\mathcal{S}=\int_{x,t} \left(\dot \fii^* \dot \fii + m^2 \fiid \fii
  +\frac{1}{\theta^2} \fiid \star x_{j} \star x_{j}\star \fii
  +\frac{1}{\theta^2} \fii \star x_{j} \star x_{j}\star \fiid \right)
  +\mathcal{S}_{int} \;,  
\fin 
where the dot denotes differentiation with respect to $x_0$. The
interaction term that we shall consider may be regarded as the orientable
analog of the $\varphi^4$ vertex, namely, 
\ecua \label{eq:vertice} 
\mathcal{S}_{int}=\frac{g}{4!}\int_{x,t} \fiid \star \fii \star \fiid
\star \fii \;.  
\fin 
Note that there is, indeed, yet another inequivalent analog to the
$\varphi^4$ vertex, namely: 
\ecua 
\mathcal{S}_{int}=\frac{g}{4!}\int_{x,t}
\fiid \star \fiid \star \fii \star \fii \;.  
\fin 
We shall not, however, deal here with a theory including this term since
its UV properties  seem to be qualitatively different~\cite{bulquen2}.
The interaction term (\ref{eq:vertice}) yields a
super-renormalizable theory, as we shall see in section
\ref{sec:renormalization}.

To carry on explicit calculations it is convenient to chose the so called
{\em matrix basis}, since, as it can be shown, their
$\star$-product adopts a `diagonal form' :
\begin{itemize}
\item $f_{nk} \star f_{k'n'}=\delta_{kk'} f_{nn'}$
\item $(f_{nk})^{*}=f_{kn}$ \;.
\end{itemize}
In Appendix A, a brief summary of this and related properties is presented. 
Careful demostrations may be found, for example, in~\cite{gracia}.

The coefficients $\varphi_{nk}(t)$, that appear in the expansion of the field
in such a basis, 
\begin{equation}
  \varphi(x,t)\;=\; \sum_{nk} \fii_{nk}(t) f_{nk}(x) \;,
\end{equation}
become then the dynamical variables.
In terms of these coefficients, the action integral reads: 
\ecua
\mathcal{S}=\int_{t_1 t_2} \fii^*_{ln}(t_1)
G_{ln,kr}(t_1-t_2)\fii_{kr}(t_2)+\frac{2\pi \theta
  g}{4!}\int_{t}\fii^{*}_{n_1,n_4}\fii_{n_1,n_2}\fii^{*}_{n_3,n_2}\fii_{n_3,n_4}\coma
\fin 
where 
\ecua 
G_{ln,kr}(t_1-t_2)=2\pi \theta
\delta(t_1-t_2) \delta_{lk}\delta_{nr}(-\partial_t^2+m^2+\frac{2}{\theta}(k+n+1))
\fin 
is a kernel that defines the quadratic (free) part of the action.  To
derive the Feynman rules corresponding to this action, we need an explicit
expression for $\Delta=G^{-1}$. Since $G$ is already diagonal with respect
to its discrete indices, we only need to deal with the temporal
coordinates. In Fourier (frequency) space:
\ecua \label{eq:propa_impulso}
\hat \Delta_{ln,kr}(\nu)=\frac{\delta_{lk}\delta_{nr}}{2 \pi \theta} \frac{1}{\omega_{nk}^2+\nu^2} \coma
\fin
and after Fourier transformation: 
\ecua
\Delta_{ln,kr}(t_1-t_2)=\braket{\fii_{ln}(t_1)\fii^*_{kr}(t_2)}_0=\frac{\delta_{lk}
\delta_{nr}}{2\pi\theta}\frac{e^{-\omega_{kn}|t_1-t_2|}}{2\omega_{kn}}
\coma 
\fin 
where
\ecua 
\omega_{kn}^2=m^2+\frac{2}{\theta}(k+n+1) \punto 
\fin 
The Feynman rules and conventions used for the diagrammatic expansion that
follows from this model are better introduced in terms of diagrams with a
double line notation, to cope with matrix indices.  Orientation is, on the
other hand, assigned according to the usual convention for creation and
annihilation operators. 

\begin{figure}[!ht]
\begin{center}
\begin{picture}(0,0)%
\includegraphics{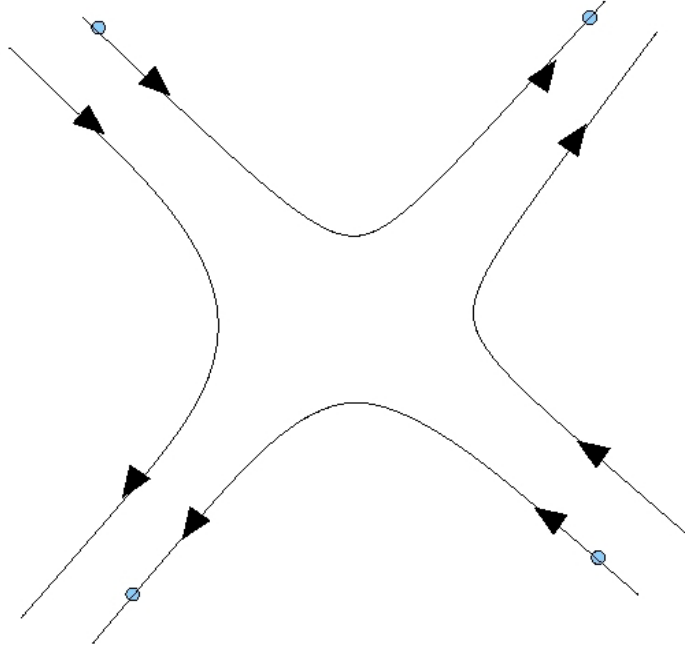}%
\end{picture}%
\setlength{\unitlength}{3108sp}%
\begingroup\makeatletter\ifx\SetFigFont\undefined%
\gdef\SetFigFont#1#2#3#4#5{%
  \reset@font\fontsize{#1}{#2pt}%
  \fontfamily{#3}\fontseries{#4}\fontshape{#5}%
  \selectfont}%
\fi\endgroup%
\begin{picture}(5580,5400)(2701,-7261)
\end{picture}%
\end{center}
\caption{The interaction vertex $\frac{g}{4!}
\fii^{*}_{n_1,n_4}\fii_{n_1,n_2}\fii^{*}_{n_3,n_2}\fii_{n_3,n_4}$} \label{fig:vertex} 
\end{figure}

The free propagator and the interaction vertex correspond to the diagrams
of figures~\ref{fig:propa} and~\ref{fig:vertex}, respectively:

\begin{figure}[!ht] 
\begin{center}	
\begin{picture}(0,0)%
\includegraphics{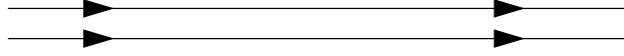}%
\end{picture}%
\setlength{\unitlength}{4144sp}%
\begingroup\makeatletter\ifx\SetFigFont\undefined%
\gdef\SetFigFont#1#2#3#4#5{%
  \reset@font\fontsize{#1}{#2pt}%
  \fontfamily{#3}\fontseries{#4}\fontshape{#5}%
  \selectfont}%
\fi\endgroup%
\begin{picture}(3714,294)(529,-208)
\end{picture}%
\end{center}
\caption{The free propagator} \label{fig:propa}
\end{figure}

A dot attached to a line indicates that it corresponds to the first index.
So when two vertices are connected with a double line, both the dots and
the orientation of the lines must coincide (note that it is not necessary
to attach a dot to the propagator).  Equipped with this notation, we may
easily group all the inequivalent diagrams corresponding to a  given class.
Symmetry factors can, of course, be calculated by standard application of
Wick's theorem.

Thus we are ready to construct perturbatively the generating functional of
1PI graphs which we shall calculate explicitly up to the one-loop order. 

This analysis is adapted for a propagator with a simple form in the matrix
basis. Other basis can be of interest (such as plane waves) depending on
the structure of the propagator and the vertex~\cite{seiberg}.
\section{Renormalization}\label{sec:renormalization} 
\subsection{One-loop divergences}
\begin{figure}[!ht]
\begin{center} 
\begin{picture}(0,0)%
\includegraphics[scale=0.6]{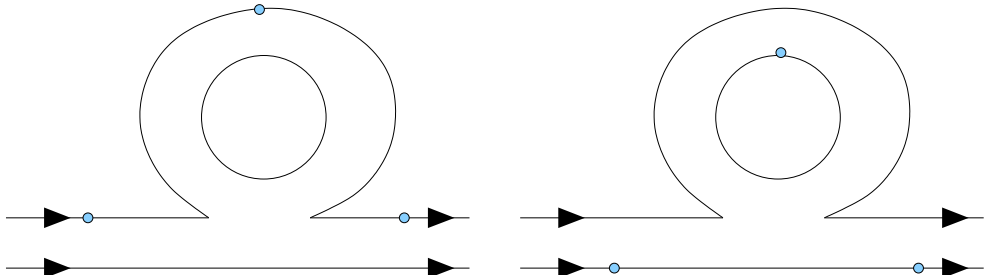}%
\end{picture}%
\setlength{\unitlength}{3947sp}%
\begingroup\makeatletter\ifx\SetFigFont\undefined%
\gdef\SetFigFont#1#2#3#4#5{%
  \reset@font\fontsize{#1}{#2pt}%
  \fontfamily{#3}\fontseries{#4}\fontshape{#5}%
  \selectfont}%
\fi\endgroup%
\begin{picture}(10149,2823)(814,-2900)
\end{picture}%
  \end{center}
  \caption{The tadpole graph. Two contractions are possible.}\label{fig:tadpole}
\end{figure}

It is easy to see
that the only divergent diagram of the theory at the one-loop level is
the~\emph{tadpole} graph of Figure~\ref{fig:tadpole}.  
 
As shown in the figure, there is a `free' internal index (not fixed by the
external ones). This leads to an UV divergent contribution to the two point
function:  
\ecua 
\Gamma_{n_0 n_1,n_2 n_3}^{(2), planar}(x-y)=\frac{2 g}{4!} \delta_{n_0 n_2} \delta_{n_1 n_3} \delta(x-y) \sum_{k\geq0}
\frac{1}{\omega_{k, n_2}} \;.
\fin
The same amplitude is obtained writing $n_3$ instead of $n_2$ and this would yield to a symmetric expression in $\varphi$ and $\varphi^*$. This corresponds to the other contraction shown in Figure \ref{fig:tadpole}. For the sake of simplicity we concentrate now in one of these, and we finally give the symmetrized expression in equation \ref{eq:contribution_two}.\\ 
An Euclidean cut off can be implemented simply by limiting the number of
modes we sum. Denoting by $k_{max}$ the maximum index in the (convergent)
sum, we split it up into two parts: one of them shall give a mass
renormalization term, while the other will be a
function of $n_2$ with a finite limit as $k_{max}\to \infty$. We chose as
subtraction point $n_2=0$, in this way the singular contribution to
$\Gamma$ is: 
\ecua 
\delta \Gamma = \frac{2 g}{\pi \theta 4!}
\Big{(}\sum_{k=0}^{k=k_{max}}\frac{1}{\omega_{k, 0}}\Big{)}\int_{x,t} \fii^*_{(x,t)} \star
\fii_{(x,t)} \;,
\fin 
which can be absorbed by the definition of the mass
parameter. 

\noindent On the other hand, the finite part reads: 
\ecua
\label{eq:finita} \sum_{k=0}^{k=k_{max}} (\frac{1}{\omega_{k,
 n_2}}-\frac{1}{\omega_{k, 0}}) \;,
\fin 
where the $k_{max}\to \infty$ limit can be taken to get a finite
contribution to the generating functional. This yields a function of $n_2$
that, as we shall see, can be written as a (one-body) potential term.

\subsection{Renormalizability and power counting}
Let us first show the theory is at least renormalizable (by power
counting). For a given Green's function the most important contributions
are given by the planar graphs. But taking into account the structure of
the propagator (\ref{eq:propa_impulso}) any amplitude must converge better
than a fermionic theory with a quartic vertex two dimensions (and without 
infrared problems). In order to show this we recall the standard definition: 
\ecua
\omega_{vertex}=(\frac{d-1}{2})F_{\nu} \;\; , 
\fin 
where $F_{\nu}$ is the number of fermions in the vertex. So in our case the
theory behaves better than $\omega_{\nu}=2$, i.e. a renormalizable
theory.

\begin{figure}[!ht] \begin{center} \begin{picture}(0,0)%
\includegraphics{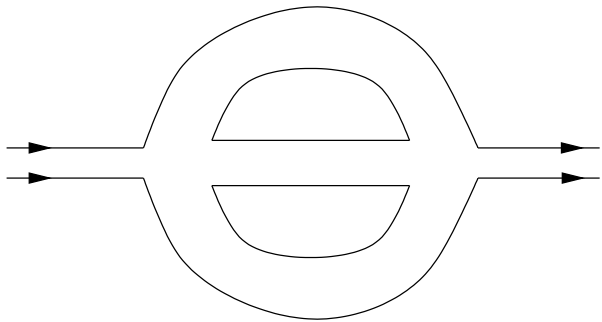}%
\end{picture}%
\setlength{\unitlength}{4144sp}%
\begingroup\makeatletter\ifx\SetFigFont\undefined%
\gdef\SetFigFont#1#2#3#4#5{%
  \reset@font\fontsize{#1}{#2pt}%
  \fontfamily{#3}\fontseries{#4}\fontshape{#5}%
  \selectfont}%
\fi\endgroup%
\begin{picture}(3534,1888)(439,-1725)
\end{picture}%
\caption{Two loop self energy diagram.} \label{fig:self_2loops}
\end{center}
\end{figure}

In order to see that the theory is super-renormalizable, note that there
must be at least two propagators in each loop (in other case, it
would be the one-loop  tadpole contribution, that has already been 
considered), but products of two or more propagators of the form 
(\ref{eq:propa_impulso}) yield
convergent integrals, because the argument can be sum or integrated in any
order and each of the iterated operation converges~\cite{folland}. One way
to see this is integrating in the worst iteration possible, this is to
perform the continuous integral and then the sum. But if one of the
propagators is multiplied by a rational function of the discrete variable
the sum converges, and this is indeed the case (as can be easy verified
performing the integral asymptotically).\\ There remain non-trivial cases,
namely: overlapping loop graph such as the one shown in Figure~\ref{fig:self_2loops}.

The amplitude associated with this diagram is proportional to:
\begin{eqnarray} \nonumber
\int_{\omega_1\omega_2}\sum_{n_1 n_2}\frac{1}{\omega_1^2+m^2+\frac{2}{\theta}(k_1+n_1+1)}\frac{1}{\omega_2^2+m^2+\frac{2}{\theta}(k_2+n_2+1)}\times\\ \frac{1}{(\omega_1+\omega_2-q)^2+m^2+\frac{2}{\theta}(n_1+n_2+1)} \; ,
\end{eqnarray}
where $k_1$, $k_2$ and $q$ are external variables. This graph is convergent iff the following integral is convergent:
\ecua
\int d^4x \frac{1}{x_1^2+|x_2|+1}\frac{1}{x_3^2+|x_4|+1}\frac{1}{(x_1+x_3-\beta)^2+|x_1|+|x_2|+1}
\coma
\fin
but this is indeed the case, because is an integral of a positive function and the integration in each variable is convergent. Any other multiloop planar diagram is convergent for the same reason. In this way we see that it is enough to renormalize the tadpole graph. 

\section{Renormalized generating functional}\label{sec:renormalizedgeneratingfunctional}
We  construct here the generating functional of 1PI graphs for the one loop
renormalized perturbation series up to fourth order in the field variable.
\subsection{Two point function}
\begin{figure}[!ht] 
  \begin{center}
\begin{picture}(0,0)%
\includegraphics{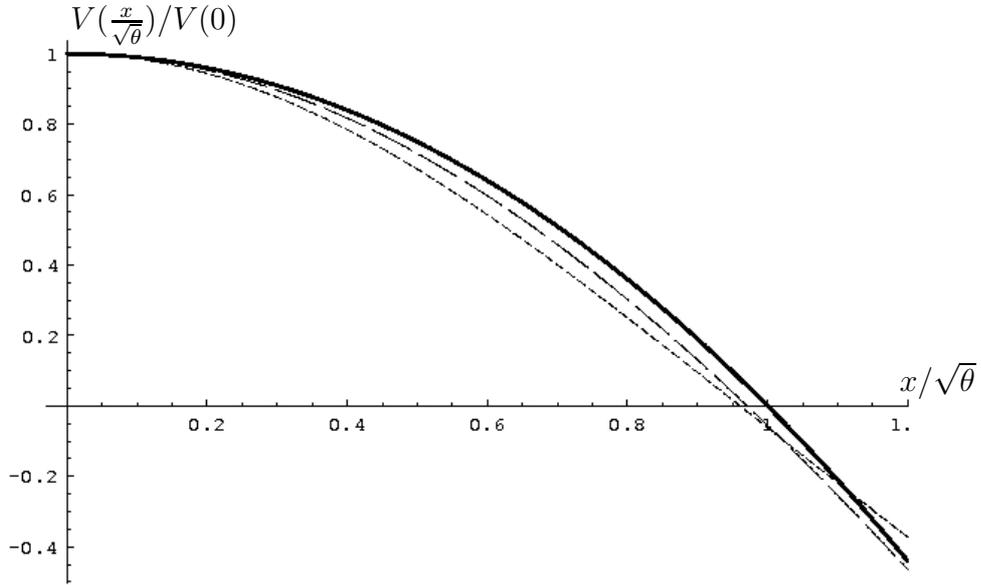}%
\end{picture}%
\setlength{\unitlength}{4144sp}%
\begingroup\makeatletter\ifx\SetFigFont\undefined%
\gdef\SetFigFont#1#2#3#4#5{%
  \reset@font\fontsize{#1}{#2pt}%
  \fontfamily{#3}\fontseries{#4}\fontshape{#5}%
  \selectfont}%
\fi\endgroup%
\begin{picture}(5625,4050)(1,-3211)
\put(451,614){\makebox(0,0)[lb]{\smash{{\SetFigFont{12}{14.4}{\rmdefault}{\mddefault}{\updefault}{\color[rgb]{0,0,0}$V(\frac{x}{\sqrt{\theta}})/V(0)$}%
}}}}
\put(5356,-1546){\makebox(0,0)[lb]{\smash{{\SetFigFont{12}{14.4}{\rmdefault}{\mddefault}{\updefault}{\color[rgb]{0,0,0}$x/\sqrt{\theta}$}%
}}}}
\end{picture}%
  \end{center}
  \caption{One-body potential due to quantum corrections. $m^2\theta=0$ (Short dashed),$m^2\theta=2$ (Long dashed), $m^2\theta\rightarrow \infty$ (bold).} \label{fig:pote}
\end{figure}

We need to consider the expression in (\ref{eq:finita}) in more detail. 
This is a convergent series which defines a holomorphic function of $n_2$.
Introducing coefficients $\alpha_{\lambda}$, so that: 
\ecua 
\sum_{k \geq0} (\frac{1}{\omega_{k,
    n_2}}-\frac{1}{\omega_{k, 0}})=\sum_{\lambda\geq1}\alpha_{\lambda}
n_{2}^{\lambda} \;, 
\fin 
the relation: 
\ecua
\alpha_{\lambda}=\sqrt{\frac{\theta}{2}}
\beta^{(\lambda)}_{(1+\frac{m^2\theta}{2})} \fin \ecua
\beta^{(\lambda)}(z)=\frac{1}{\lambda!}\frac{\partial^{\lambda}}{\partial
  w^{\lambda}}\Big{(}\mathcal{Z}[\frac{1}{2},w+z]-\mathcal{Z}[\frac{1}{2},z]\Big{)}|_{w=0}
\fin 
where $\mathcal{Z}$ is the Hurwitz zeta function, is easily obtained. 
It is important to note the smooth behavior with respect to the product
$m^2\theta$, this number is greater than zero so the argument of the
function beta is always greater than
one (i.e. in this domain the function is regular). Now we are ready to
include the contribution of the two point function to the generating
functional. The singular part is absorbed in a mass renormalization, while 
the finite part is: 
\ecua 
\Gamma_{n_0 n_1,n_2
  n_3}^{(2), finite}(x-y)=\frac{2 g}{4!} \delta_{n_0 n_2}
\delta_{n_1 n_3} \delta(x-y) \left(\sum_{\lambda\geq1}\alpha_{\lambda}
n_{2}^{\lambda}\right)\;. 
\fin 
Taking now into account the correspondence with the functional
representation (see Appendix A), we can use the number operator to get an
expression in the original functional space:  
\ecua \label{eq:contribution_two}
\delta \Gamma[\fii,\fii^*]=\frac{2 g}{4!2\pi \sqrt{2\theta}}\int_{x,t}\big{(}\fii^* \star
V(\frac{x}{\sqrt{\theta}}) \star \fii + \fii \star V(\frac{x}{\sqrt{\theta}})
\star \fii^* \big{)}\;, 
\fin 
where we have used the definition: 
\ecua
V(\frac{x}{\sqrt{\theta}}) \,=\,
\sum_{\lambda\geq1}\frac{\beta^{(\lambda)}}{2^\lambda}\big{(}\frac{x_{j}
\star x_{j}}{\theta}-1\big{)}^{\star \lambda} \;. 
\fin 
This shows the explicit  form of the one-body potential,
The first three terms in the expansion of this potential are plotted in
Figure \ref{fig:pote}, for the values $m^2\theta = 0$, $m^2\theta = 2$ and $m^2\theta = \infty$ .

It is clear that this quantum correction tends to deconfine the system, as
it should be expected from the repulsive character of the interaction.
\subsection{Four-point functions}
Now we deal with the four-point contributions, which correspond to four
inequivalent diagrams, which we study below, together with their
corresponding contributions to the action.  
\begin{figure}[!ht]
  \begin{center}
    \begin{picture}(0,0)%
      \includegraphics{4ptosA.pstex}%
    \end{picture}%
    \setlength{\unitlength}{3108sp}%
    \begingroup\makeatletter\ifx\SetFigFont\undefined%
    \gdef\SetFigFont#1#2#3#4#5{%
      \reset@font\fontsize{#1}{#2pt}%
      \fontfamily{#3}\fontseries{#4}\fontshape{#5}%
      \selectfont}%
    \fi\endgroup%
    \begin{picture}(6300,2700)(3151,-4201)
    \end{picture}%
  \end{center}
  \caption{} \label{fig:4A}
\end{figure}
The diagram of Figure \ref{fig:4A} contributes with: 
\ecua 
\delta \Gamma=- \frac{S(2\pi \theta
  g)^2}{2(4!)^2}\delta_{(t_1-t_2)}\delta_{(t_3-t_4)}\delta^{n_1}_{n_3}
\delta^{n_2}_{n_0}\delta^{n_5}_{n_7}\delta^{n_4}_{n_6}
(\Delta^{n_4 n_1,n_4 n_1}_{(t_1-t_3)})^2 
\coma
\fin 
where $S$ is a symmetry factor.
To obtain an explicit expression for the quantum correction to the action we
will consider a low energy approximation, assuming we are concerned with the 
physics of this system up to $n_i^{max}$ (which is a kind of low-momentum
approximation). 

\noindent Thus, assuming the condition $\theta m^2 >>
n_i^{max}$ for the external indices, we can write: 
\ecua \label{eq:np1}
\delta \Gamma= - \alpha
\frac{g^2}{m^3 \theta^2} \int_t (\int_x \fii^*_{(x,t)} \star
\fii_{(x,t)})^2 \; , \;\;\;\alpha>0 
\coma
\fin 
where $\alpha$ is independent of the parameters of the problem.
Another contribution is the one represented in Figure \ref{fig:4B}.
\begin{figure}[!ht] 
  \begin{center}
    \begin{picture}(0,0)%
      \includegraphics{4ptosB.pstex}%
    \end{picture}%
    \setlength{\unitlength}{3108sp}%
    \begingroup\makeatletter\ifx\SetFigFont\undefined%
    \gdef\SetFigFont#1#2#3#4#5{%
      \reset@font\fontsize{#1}{#2pt}%
      \fontfamily{#3}\fontseries{#4}\fontshape{#5}%
      \selectfont}%
    \fi\endgroup%
    \begin{picture}(3600,2700)(3601,-5101)
    \end{picture}%
  \end{center}
  \caption{}\label{fig:4B}
\end{figure}
Its analytic expression is: 
\ecua 
\delta \Gamma=\frac{-S(2\pi \theta
  g)^2}{2(4!)^2}\delta_{(t_1-t_3)}\delta_{(t_2-t_4)}\delta^{n_6}_{n_4}\delta^{n_3}_{n_1}\delta^{n_2}_{n_0}\delta^{n_5}_{n_7}
\Delta^{n_1 n_4,n_1 n_4}_{(t_2-t_3)} \Delta^{n_0 n_5,n_0 n_5}_{(t_2-t_3)} 
\punto
\fin
Using the same approximation as for the previous diagram, we see that it
may be approximated by 
\ecua\label{eq:np2}
\delta \Gamma= - \alpha \frac{g^2}{m^3 \theta^2} \int_t (\int_x \fii^*_{(x,t)}
\star \fii_{(x,t)})^2 \;\;,\;\;\;\alpha>0 \;. 
\fin
Another nonequivalent diagram of this class is represented in Figure \ref{fig:4C}.
\begin{figure}[!ht] 
  \begin{center}
    \begin{picture}(0,0)%
      \includegraphics{4ptosC.pstex}%
    \end{picture}%
    \setlength{\unitlength}{3108sp}%
    \begingroup\makeatletter\ifx\SetFigFont\undefined%
    \gdef\SetFigFont#1#2#3#4#5{%
      \reset@font\fontsize{#1}{#2pt}%
      \fontfamily{#3}\fontseries{#4}\fontshape{#5}%
      \selectfont}%
    \fi\endgroup%
    \begin{picture}(3600,2880)(4276,-5101)
    \end{picture}%
  \end{center}
  \caption{}\label{fig:4C}
\end{figure}
Under the same approximation we used before, it contributes with: 
\ecua \label{eq:np3}
\delta \Gamma= - \alpha \frac{g^2}{m^3 \theta^2} \int_t (\int_x \fii^*_{(x,t)}
\star \fii_{(x,t)})^2 \;\;,\;\;\;\alpha>0 \;. 
\fin
Thus under this approximation all non-planar contributions have the same expression. Numerical factors (we call $\alpha$ in equations \ref{eq:np1}, \ref{eq:np2} and \ref{eq:np3}) can of course be different.\\
There is also a planar diagram with one of its indexes not fixed by the
external ones, Figure \ref{fig:4D}. 
\begin{figure}[!ht]
  \begin{center}
    \begin{picture}(0,0)%
      \includegraphics{4ptosD.pstex}%
    \end{picture}%
    \setlength{\unitlength}{3108sp}%
    \begingroup\makeatletter\ifx\SetFigFont\undefined%
    \gdef\SetFigFont#1#2#3#4#5{%
      \reset@font\fontsize{#1}{#2pt}%
      \fontfamily{#3}\fontseries{#4}\fontshape{#5}%
      \selectfont}%
    \fi\endgroup%
    \begin{picture}(5760,3150)(3601,-5461)
    \end{picture}%
  \end{center}
  \caption{} \label{fig:4D}
\end{figure}
Because of this its contribution is more important than the previous ones:
\ecua 
\delta \Gamma=\frac{-S(2\pi \theta
  g)^2}{2(4!)^2}\delta_{(t_1-t_2)}\delta_{(t_3-t_4)}\delta^{n_0}_{n_2}\delta^{n_6}_{n_4}\delta^{n_3}_{n_5}\delta^{n_1}_{n_7}
\sum_{\lambda \geq 0} \Delta^{\lambda n_3,\lambda n_3}_{(t_1-t_3)} \Delta^{\lambda
  n_1,\lambda n_1}_{(t_1-t_3)} 
\coma
\fin 
which we again approximate, with the result: 
\ecua 
\delta \Gamma= - \alpha
(g\theta^{\frac{1}{2}})\mathcal{Z}_{(\frac{3}{2},\frac{2+\theta m^2}{2})} g
\int \fii^* \star \fii \star \fii^* \star \fii \;\; , \;\;\;\alpha>0 
\punto
\fin 
This diagram has a finite $\theta m^2 \rightarrow \infty$ limit.
Indeed,  
\ecua 
\lim_{x \to \infty} \sqrt{x}
\mathcal{Z}_{(\frac{3}{2},\frac{2+x}{2})}=\beta 
\coma
\fin 
where $\beta$ is a positive number of order unity. 
So we have 
\ecua 
\delta \Gamma= - \alpha \beta
\frac{g^2}{m}\int \fii^* \star \fii \star \fii^* 
\star \fii \;\; , \;\;\; \alpha>0 \;.
\fin 
In this way, we see that only the last graph is leading when
$\theta m^2 \rightarrow \infty$. This is a consequence of the free internal
line (loop) which gives the most important contribution to the generating 
functional in this limit.
\subsection{Approximate generating functional}
Joining all the previous pieces, we get an approximate expression for the 
1PI functional, in the $\theta m^2 \rightarrow \infty$ limit.  
\ecua \nonumber
\Gamma=\mathcal{S}+\frac{g}{4!2\pi \sqrt{2\theta}}\int_{x,t}\fii^* \star
V(\frac{x}{\sqrt{\theta}}) \star \fii + \fii \star V(\frac{x}{\sqrt{\theta}})
\star \fii^*- \fin \ecua -\alpha \beta \frac{g^2}{m}\int \fii^* \star \fii
\star \fii^* \star \fii 
\punto
\fin
The approximation have been used to eliminate some of the four point
contributions. Taking into account the form of the coefficients in the two
point function it is easily verified that if the series which defines the
one-body potential is truncated, this correction vanishes as well. We do
not have, however, a closed analytical expression for that correction, so this term should be kept.\\
If further corrections are taken into account under the approximation $\theta m^2 \rightarrow \infty$ non-planar diagrams can be eliminated as in the four point function case. It is easily seen that if a series of internal lines connected to external legs are replaced by an internal loop the amplitude results a factor $\theta m^2$ bigger than the non-planar case. So, for example, if the two-loop self energy diagram is considered as in Figure \ref{fig:self_2loops} the non-planar case would be suppressed by a factor $\mathcal{O}(\frac{1}{(\theta m^2)^2})$, and so the latter correction would not be important.

\section{Non trivial vacuum configurations}\label{ntvc} 
Using the properties of the matrix base, exact classical solutions to the
equations of motion can be found. A natural question is whether we
can define a sensible quantum theory around those non trivial vacuum
configurations.  As we shall see, this is indeed the case. 
We will also analyze how the vacuum energy is shifted under variations of
the parameters that characterize the solutions.
\subsection{Classical solutions}
Considering the real-time action associated to the Euclidean one of
(\ref{eq: accion}), we see that a classical solution must satisfy: 
\ecua 
\ddot \fii + m^2 \fii + \frac{1}{\theta^2}(x_{\mu}\star x_{\mu}\star
\fii)+\frac{1}{\theta^2}(\fii \star x_{\mu}\star x_{\mu})+\frac{2g}{4!} \fii
\star \fiid \star \fii = 0 \;. 
\fin 
Using the ansatz 
\ecua\label{ansatz}
\varphi_{nk}(x,t)=e^{\im \Omega_{nk} t} f_{nk}(x) 
\coma
\fin
we have a solution to the nonlinear problem if the following dispersion 
relation is satisfied: 
\ecua
\Omega_{nk}^2=m^2+\frac{2}{\theta}(n+k+1)+\frac{2g}{4!} 
\punto
\fin
This means that, at the classical level, objects with typical size $\theta$ 
can be stable (note the difference with the commutative case). There is a vast literature on the subject of solitonic solutions for noncommutative theories, some basic references are \cite{nekrasov} and \cite{Schaposnik:2004rc}.\\  
In order to study the quantum corrections, we deal next with 
the Euclidean version of the problem.

\subsection{Quantum case}
Consider again the Euclidean action (\ref{eq: accion}). The condition for an extremum
with an ansatz such as (\ref{ansatz}) is
\ecua
\Omega_{nk}^2+m^2+\frac{2}{\theta}(n+k+1)+\frac{2g}{4!}=0 \;.
\fin 
If we focus on time-independent solutions, a symmetry-breaking like
potential is needed in order to have an extremum. We will, however, continue
the discussion for a different kind of solution. As it may be easily verified,
$\varphi=\eta f_{00}$ is a solution of the equation of motion if: 
\ecua \label{eq:condition}
m^2+\frac{2}{\theta}+\frac{2g\eta^2}{4!}=0
\punto
\fin 
In the same way it is possible to generate more solutions of the form $\varphi=\eta f_{nk}$, with a non-linear condition for the amplitude. We will focus on the fundamental one ($\varphi=\eta f_{00}$) for an explicit analysis.\\
A first question is whether a generating functional (in the path
integral formalism) can be constructed by expanding around this extremum. Next we want to know the dependence
of the vacuum energy with the parameters of the problem. Let us first deal with
the first (stability)  condition. The second-order correction about the 
extremum of the Euclidean action is parameterized as follows: 
\ecua 
\frac{1}{2} \left(
  \begin{array}{ll}
    \chi \\ 
    \chi^* 
  \end{array} 
\right)^{\dagger} \mathbb{H}(\mathcal{S}) \left(
  \begin{array}{ll}
    \chi \\ 
    \chi^* 
  \end{array} 
\right) 
\fin 
where $\chi$ is the fluctuation around the non trivial solution,
and $\mathbb{H}(\mathcal{S})$ is the Hessian matrix:
\ecua 
\left(
  \begin{array}{ll}
    \frac{\delta^2 \mathcal{S}}{\delta \varphi_1 \delta \varphi^*_2} & 
    \frac{\delta^2 \mathcal{S}}{\delta \varphi^*_1 \delta
      \varphi^*_2} \\\frac{\delta^2 \mathcal{S}}{\delta \varphi_1 \delta
      \varphi_2} & \frac{\delta^2 \mathcal{S}}{\delta \varphi^*_1 \delta
      \varphi_2} 
  \end{array} 
\right) 
\fin
with the usual notation for kernels.  So the consistency condition is
equivalent to check that all eigenvalues of this matrix are positive. 
In fact we already have a basis of eigenvectors $\{\varphi/\varphi=e^{i\Omega
  t}f_{nk}(x),~n,k \in \mathbb{N},~\Omega \in \mathbb{R}\}$, 
and the eigenvalues are: \begin{displaymath} \left\{
    \begin{array}{ll} \Omega^2+m^2+\frac{2}{\theta}(n_1+n_2+1) & n_1,n_2 \geq 1
      \\ \\ \Omega^2+m^2+\frac{2}{\theta}(n_1+n_2+1)+\frac{2g\eta^2}{4!} &
      n_{1,2}=0,  n_{2,1} \geq 1\\
      \\\Omega^2+m^2+\frac{2}{\theta}+\frac{6g\eta^2}{4!} & n_1=n_2=0
    \end{array}\right.
\end{displaymath}
Using the condition \ref{eq:condition} the set of eigenvalues is:
\begin{displaymath} \left\{
    \begin{array}{ll} \Omega^2+\frac{2}{\theta}(n_1+n_2)-\frac{2g\eta^2}{4!} & n_1,n_2 \geq 1
      \\ \\ \Omega^2+\frac{2}{\theta}(n_1+n_2) &
      n_{1,2}=0,  n_{2,1} \geq 1\\
      \\\Omega^2+\frac{g\eta^2}{3!} & n_1=n_2=0,
    \end{array}\right.
\end{displaymath}
which are all positive if $g\eta^2<\frac{2\;4!}{\theta}$.\\
Now the vacuum energy shift between two sets of parameters associated with the
eigenvalues $\{\lambda'_n(\Omega)\}$ and $\{\lambda_n(\Omega)\}$ can be
evaluated as: 
\ecua 
\Delta E=\frac{1}{2\pi}\int_{d\Omega}
\sum_{n}\log \Big{(}\frac{\lambda'_n(\Omega)}{\lambda_n(\Omega)}\Big{)} \punto
\fin 
This shows there is a way of changing the parameters such that the energy remains constant, if we mantain $m$ and $\theta$ constant and if the product $g\eta^2$ does not change then $\Delta E=0$. But note that we can change the coupling constant $g$ and the amplitude of the solution $\eta$, with just one constraint.

In reference~\cite{0709}, a  throughout study of non-trivial vacuum
configurations in (real and complex) scalar models
in  $2$ and  $4$ spacetime dimensions is presented.
The kind of ansatz that we consider here may be regarded as an embedding
to $2+1$ dimensions, of one of the solutions considered there for the
$2$-dimensional case.

\section{Conclusions}\label{sec:conclusions}
We have shown explicitly that the self-dual model is a super-renormalizable 
theory, carrying out the explicit one-loop renormalization procedure, and
evaluating the corresponding contributions to the effective action to that
order. We have also found an approximate expression for the generating 
functional of proper vertices, under the assumption: $m^2\theta>>1$.  

Besides, some non trivial solutions in the presence of the GW term and a
symmetry breaking potential have been found at classical level, and it was
shown that they are stable under the leading quantum corrections, by
evaluating the exact eigenvalues of the Hessian around those extrema. The
resulting dependence of the vacuum energy on the model's parameters has
also been explicitly found.
\section*{Acknowledgments} 
C.~D.~F. was supported by CONICET, ANPCyT and
UNCuyo. G.~A.~M. is supported by a Petroenergy SA - Trafigura studentship at Instituto Balseiro UNCuyo. 

We thank Professors A. de Goursac, A. Tanasa and J-C Wallet for pointing
out reference~\cite{0709}, and the relation between our classical
configuration and some non-trivial vacuum configurations considered
there, for even-dimensional models.

\section*{Appendix A}
In this section we will briefly derive the properties of the basis which `diagonalizes' Moyal product: \ecua (f \star g)_{(x)}=f_{(x)} e^{\frac{\im}{2}
  \theta_{\mu \nu}\overleftarrow{\partial_\mu} \overrightarrow{\partial_\nu }}
g_{(x)} \;\; .\fin First we build an operatorial representation of the algebra.
Consider two hermitian operators such that $[x_1, x_2]=i \theta$, and
define the creation an annihilation operators $a$ and $a^{\dagger}$: \ecua
a=\frac{x_1 + i x_2}{\sqrt{2\theta}} \hspace{8mm} [a,a^\dagger]=1 \;\; .
\fin To connect the algebra of functions $\mathcal{G}$ with the algebra of
operators $\mathcal{G}'$ consider a map $\mathcal{S}^{-1}:
f\in\mathcal{G}\rightarrow \mathcal{O}_{(f)}\in \mathcal{G}'$ defined by the
following equation (here time is just a parameter): \ecua \label{eq: symbol}
\mathcal{O}_f(t)=\int_{\bar k\in \mathbb{R}^2}\frac{1}{(2\pi)^2}\hat f(\bar
k,t):e^{i \sqrt{\frac{\theta}{2}}(k^* a+ka^\dagger)}: \hspace{4mm}; k=k_1+i
k_2 \coma \fin where $::$ denotes normal ordering and $\hat f$ is the usual Fourier
transform: \ecua \hat f(\bar k,t)=\int_{\bar x\in \mathbb{R}^2} f(\bar x,t)
e^{-i \bar k_j \bar x_j} \punto \fin It is the work of a moment to verify the properties:
\begin{itemize}
\item \ecua \label{eq: Moyal_ope} \mathcal{O}_{f\star
    g}=\mathcal{O}_f\mathcal{O}_g \fin
\item \ecua Tr(\mathcal{O}_f)=\frac{1}{2\pi \theta}\int_{x} f(x, t)
  \fin
\item \ecua \mathcal{O}_{(\partial_{x_1} f)}=[\frac{\im}{\theta} x_2,
  \mathcal{O}_f] \hspace{4mm} \mathcal{O}_{(\partial_{x_2}
    f)}=[-\frac{\im}{\theta} x_1, \mathcal{O}_f] \;\; .\fin
\end{itemize}
So the Moyal product in the algebra of functions is mapped to composition of
operators. On the other hand, there is a special class of operators that allow
a very easy way to perform composition, namely the ones which have the form
$\ket{n}\bra{k}$. So if we know $f_{nk}\in
\mathcal{G}/\mathcal{O}_{(f_{nk})}=\ket{n}\bra{k}$ we would have \ecua f_{nk}
\star f_{k'n'}=\delta_{kk'} f_{nn'} \hspace{4mm} f_{nk}^{*}=f_{kn} \; , \fin
because of equation (\ref{eq: Moyal_ope}). This is the basis we mentioned above. To get an explicit form of the condition
$\mathcal{O}_{f_{nk}}=\ket{n}\bra{k}$ is enough to take matrix elements in equation
(\ref{eq: symbol}) and use that Laguerre associated polynomials ($\mathbb{L}^{(n-j)}_j$) are complete
(very useful identities can be found in \cite{gango}). Looking at the
coefficients we find that the Fourier transform of such a function in
polar coordinates is: \ecua \hat f_{nj(\rho,\varphi)}=2\pi\theta
\sqrt{\frac{j!}{n!}}(i \sqrt{\frac{\theta}{2}})^{j-n} e^{-i \varphi (n-j)}
\rho^{j-n} e^{-\frac{\theta \rho^2}{4}} \mathbb{L}^{(n-j)}_j(\frac{\theta
  \rho^2}{2}) \;\; ,\fin so for example a diagonal one is a Gaussian times a
polynomial \ecua \hat f_{nn(k)}=2\pi\theta \; e^{-\frac{\theta k^2}{4}}
\; \mathbb{L}^{(n)}\big{(}\frac{\theta k^2}{2}\big{)} \;\; . \fin

\end{document}